\documentclass[letterpaper, 10 pt, conference]{ieeeconf}  

\IEEEoverridecommandlockouts                              

\usepackage{graphicx} 
\usepackage{array}
\usepackage{arydshln}

\newenvironment{boxed}
    {
    \medskip
    \begin{center}
    \begin{tabular}{|p{0.9\linewidth}|}
    \hline
    }
    { 
    \\\hline
    \end{tabular} 
    \end{center}
    \medskip
    }

\title{ \LARGE \bf Human-Robot Co-Creativity: A Scoping Review \\
\large \vspace{0.5em} Informing a Research Agenda for Human-Robot Co-Creativity with Older Adults}

\author{Marianne Bossema$^{1,2}$, Somaya Ben Allouch$^{3,4}$, Aske Plaat$^{2}$ and Rob Saunders$^{2}$
\thanks{$^{1}$ Amsterdam University of Applied Sciences, The Netherlands {\tt\small m.bossema@hva.nl}}%
\thanks{$^{2}$ Leiden University, The Netherlands}%
\thanks{$^{3}$ Amsterdam University of Applied Sciences, The Netherlands}%
\thanks{$^{4}$ University of Amsterdam, The Netherlands}%
}

\begin{document}

\maketitle
\thispagestyle{empty}
\pagestyle{empty}

\begin{abstract}

This review is the first step in a long-term research project exploring how social robotics and AI-generated content can contribute to the creative experiences of older adults, with a focus on collaborative drawing and painting. We systematically searched and selected literature on human-robot co-creativity, and analyzed articles to identify methods and strategies for researching co-creative robotics. We found that none of the studies involved older adults, which shows the gap in the literature for this often involved participant group in robotics research. The analyzed literature provides valuable insights into the design of human-robot co-creativity and informs a research agenda to further investigate the topic with older adults. We argue that future research should focus on ecological and developmental perspectives on creativity, on how system behavior can be aligned with the values of older adults, and on the system structures that support this best. 
\end{abstract}

\section{INTRODUCTION}

The world's population is rapidly aging. According to the United Nations, the number of people aged 60 years or older is expected to more than double by 2050, reaching approximately 2.1 billion \cite{united2022world}. This demographic shift is having significant social, economic, and health implications. In 2019, the World Health Organization published a review of 900 studies, concluding that creative activities can promote health and well-being and help prevent and slow age-related physical and cognitive decline \cite{fancourt2019evidence}. Here, the term `creative activities' is referring to forms of personal, everyday creativity, such as making music, drawing, dancing, or crafts. According to Cohen~\cite{cohen2005mature}, such acts of everyday creativity are fundamental to psychological development and well-being in later life.

Creech et al.~\cite{creech2020creativity} present a systematic literature review into creativity and the quality of later life, which highlights the benefits of the collaborative and relational nature of creativity. Co-creativity is also linked to well-being; Zeilig et al.~\cite{zeilig2019co} suggest that sharing agency in co-creative activities can empower people with dementia. These studies share the view that co-creativity can foster social connections and create a safe space that facilitates involvement and sharing.

Social robots are playing a growing role in healthcare and well-being~\cite{broekens-heerink-rosendal:2009assistive}~\cite{allouch2020social}. There are few examples, however, of creative robot applications for older adults. Social robots offer unique opportunities to support creativity through assistance and social interaction. In addition, technological advancements in generative AI bring new opportunities to suggest tailored content in creative collaborations. There are unanswered questions, however, on how to design appropriate human-robot co-creative systems. A scoping review was conducted, to systematically map the research done in Human-Robot Co-Creativity (HRCC), and to inform a research agenda for HRCC with older adults. We define HRCC as ``An interactive system for collaborative creativity between human and robot. It involves the joint effort of two or more embodied and co-present agents, where both take initiative in response to the other(s) and contribute to creative outcomes''. This definition is based on the concepts of ``Humbots'', as described by Lubart et al., \cite{lubart_creativity_2021} and ``Mixed-Initiative Creative Interfaces'' as introduced by Deterding et al. \cite{deterding2017mixed}.

In the next section, we begin by presenting theories related to the value-sensitive design of HRCC for older adults. In Section~\ref{sec:method}, we describe the methodology we used for this scoping review. Based on the analysis of selected articles, we document the results in Section~\ref{sec:results}, followed by a discussion in Section~\ref{sec:discussion}. In Section~\ref{sec:conclusion} we provide a conclusion, leading to a research agenda outlined in Section~\ref{sec:agenda}.

\section{BACKGROUND}\label{sec:background}

Investigating the prospective role of social robots in co-creative systems bridges the fields of Human-Robot Interaction, Computational Creativity, and Arts \& Health. Here, we present theories from these fields in the context of value-sensitive design of HRCC for older adults.

\subsection{Creativity and Values of Older Adults}

Definitions of creativity generally share the common theme that it involves generating something new, valuable, and surprising~\cite{runco-jaeger:2012standard}. There are different approaches, however, to understanding this complex concept. Gl{\u{a}}veanu~\cite{gluaveanu2013rewriting} takes an ecological perspective, describing creativity as a phenomenon that emerges through interaction in a social and material environment. Kaufman \& Beghetto~\cite{kaufman2009beyond} take a developmental perspective, looking at the individual. Individuals are more likely to be creative when they are given challenging tasks that require new solutions, have a degree of autonomy and control over their work, and can collaborate and communicate effectively with others.

\begin{figure}[tb]
\centering
\includegraphics[width=0.7\linewidth]{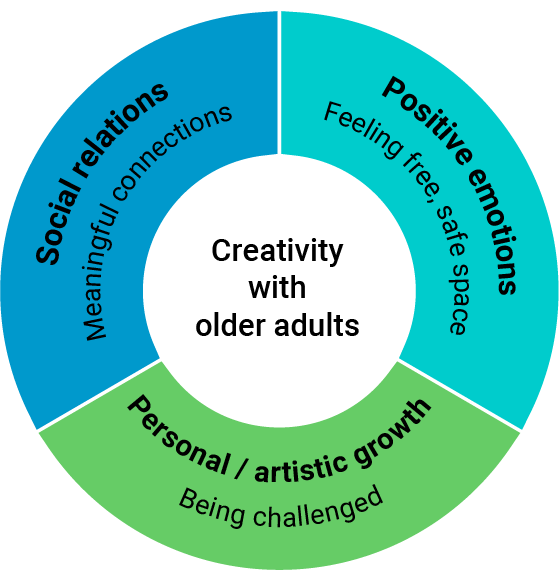}
\caption{Values that older adults attribute to creative activities, based on a Dutch nationwide study by Groot et al.~\cite{groot2021value}. Figure adapted with permission.} 
\label{fig:values}
\end{figure}

Both ecological and developmental perspectives align with the values that older adults attribute to their creative experiences. In a Dutch study by Groot et al.~\cite{groot2021value}, older participants reported appreciating creative activities for 1) offering an environment where they feel safe, accepted, and free, 2) promoting personal and artistic growth, and 3) enabling meaningful social interactions (Fig.~\ref{fig:values}). Based on the study of Groot et al., Liu et al.~\cite{liu2022participatory} investigate the relationship between context, mechanisms, and outcomes, and mention `a welcoming environment' as a consistent underlying mechanism. Liu et al.~\cite{liu2022participatory} recommend deepening our understanding of environments and affective atmospheres in art activities with older adults. Groot et al.~\cite{groot2021value} recommend Participatory Action Design as a research approach to capture the essence of older participants' creative experiences.

In ``A Roadmap for Therapeutic Computational Creativity'', Pease et al.~\cite{pease-et-al.:2022roadmap} delve into the connection between Computational Creativity and mental health and well-being. The authors discuss the benefits and risks associated with this connection. They also highlight potential opportunities, such as \emph{casual creators}~\cite{compton2015casual}. Casual creators prioritize the pleasure of the creative process over the end product and offer enjoyable and easily accessible creative experiences that may be valuable for older adults. This kind of experiences may promote ``personal and artistic growth'', while contributing to ``an atmosphere in which people feel safe, accepted, and free'' ~\cite{groot2021value}. The roadmap also discusses the concept of the `third hand', a metaphor for the therapist's role in supporting and encouraging patients' creative processes, without imposing their own ideas or disrupting the patient's autonomy~\cite{kramer1986art}. The researchers recommend collaboration with health professionals, to determine the limitations and possibilities of therapeutic computational creativity~\cite{pease2022roadmap}.

\subsection{Interaction Design for Computational Creativity}

Kaufman \& Beghetto \cite{kaufman2009beyond} mention two main requirements for people to be creative 1) a degree of autonomy and control, and 2) effective communication with others. These requirements present challenges for the design of co-creative systems. While traditional creativity support tools focus on human control, Gemeinboeck \& Saunders~\cite{gemeinboeck2016performance} suggest embodied creative agents that share the world with humans, and act autonomously, beyond their creator's intent. Mixed-Initiative Creative Interfaces~\cite{deterding2017mixed} are in between, a form of AI-enabled Creativity Support tools, where both humans and the system can take initiative during creative collaboration. This raises questions, e.g., on how agency can be shared and how initiative can be negotiated to support both well-being and mutual creativity. 

The requirement of effective communication also poses interaction design challenges. Bray \& Bown~\cite{bray_applying_2016} argue that computational creativity systems are often complex and opaque, limiting visibility and clarity of their conceptual models. Understanding may be improved when users can clearly perceive the system's structure, and develop a mental model of how this structure leads to behavior. This is crucial to facilitate a suitable level of autonomy and control. Dialogues can be expected to contribute to understanding and common ground, either language-based or through creative artifacts. A dialogic approach, as suggested by Bown et al.~\cite{bown2020speculative}, can enable both human and artificial agents (e.g. social robots) to actively influence the creative process and products, and adapt to the other's behavior.

Social robots offer unique opportunities for embodied interaction, sharing agency, and (non-)verbal communication. They can suggest tailored AI-generated content and support creative exploration. The articles being reviewed shed light on how interaction design challenges may be faced and how solutions may be applied.

\begin{table}[h]
\renewcommand{\arraystretch}{1.2}
\caption{ Keywords searched in databases}
\label{tab:search-keywords}
\begin{tabular}{|c|c|c|}
\hline
Actors & Activities & Applications\\
\hline
Human-machine & Co-creativ* & Creativ* support\\
Human-computer & Creative collaboration & Support* creativity\\
Human-robot & Art* collaboration & Stimulat* creativity\\
Human-AI & Collaborative creativity & Art therapy\\
Robot* & Collaborative art & Creativ* Robot*\\
Artificial Intelligence & Collaborative drawing & \\
AI & Collaborative painting & \\
Machine Learning & Collaborative sketching & \\
\hline
\end{tabular}
\end{table}

\section{METHOD}\label{sec:method}
Six databases were used to conduct the scoping review: ACM, IEEE, Google Scholar, PsycINFO, Pubmed, and Scopus. Keywords were chosen for \textit{`Actors' }(e.g. human-robot), \textit{`Activities'} (e.g. co-creativity), and \textit{`Application'} (e.g. Creativity Support), see Table~\ref{tab:search-keywords}. Only conference and journal articles, published in English were included. The search results (n=827) were collected in February 2023 and imported in Rayyan~\cite{ouzzani2016rayyan}, where duplicates (n=100) were removed, and labels were assigned. Searching and selecting articles was done systematically using PRISMA guidelines ~\cite{moher2009preferred}, currently with a single reviewer (the first author). Based on a first screening of titles and abstracts, articles (n=432) were excluded when they a) did not involve human subjects in evaluating a robotic system, b) described a distinctive context (e.g., business, innovation, teaching, or product development) when they were conference workshop calls or proposals, or c) were found to be duplicates. In a second screening, papers (n=157) were removed that did not show evidence of a `co-creative agent', here defined as a computational actor involved in building shared creative artifacts, in co-presence with one or more human collaborators~\cite{lubart_creativity_2021}. In the last step of the screening, studies that did not involve an embodied intelligent agent were removed (n=113). Doing the exclusion selection in separate steps allowed for acquiring a broader view, and offered the opportunity to also keep studies with non-robotic agents in mind. Two papers were added through forward and backward citation searches, and a final set of 27 articles was used for analysis.  

\begin{figure}[t]
\centering
\includegraphics[width=\linewidth]{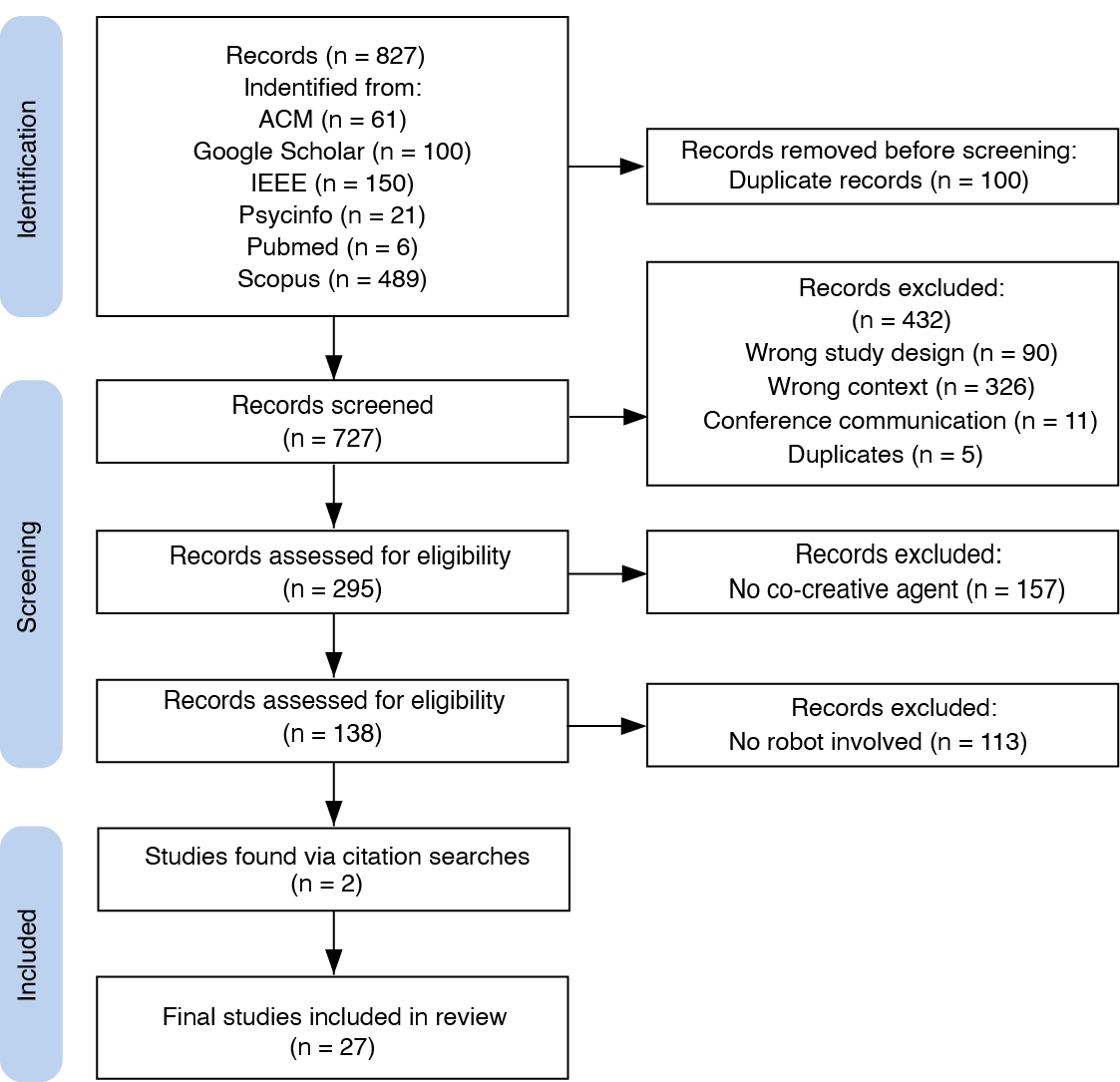}
\caption{PRISMA flow diagram for the scoping review.}
\label{fig:prisma}
\end{figure}

\subsection{Analysis of Design Research}

Studies in HRCC are a form of design research, focused on understanding specific interaction design problems. We used the Function-Behavior-Structure (FBS) ontology, as described by Gero \& Kannengiesser~\cite{gero2014function}, for analyzing and comparing this kind of study. The ontology is based on the notion that all designs can be represented in a uniform way, and that design systems can be conceptualized in three ontological categories. Function (F) is about `what the system is for', Behavior (B) covers `what it does', and Structure (S) describes the components and their relationships, or `what it consists of'. In addition, we applied a layered framework for interactions between creative collaborators, proposed by Kantosalo et al.~\cite{kantosalo2020modalities}, to the `Behavior' of co-creative systems into interaction layers of modalities, styles, and strategies to provide a finer-grained view of `what a co-creative system does'.

\begin{table*}[t]
\renewcommand{\arraystretch}{1.6}
\caption{Reviewed papers, structured using the FBS ontology~\cite{gero2014function} and the \\
Interaction Framework for Human-Computer Co-Creativity~\cite{kantosalo2020modalities}.}
\label{tab:review}
\centering
\setlength\dashlinedash{0.3pt}
\setlength\dashlinegap{2pt}
\resizebox{\textwidth}{!}{\begin{tabular}{*{9}{|c}|}
\hline
\multicolumn{3}{|c|}{Function} & \multicolumn{4}{|c|}{Behavior} & Structure & \\
\multicolumn{3}{|c|}{ } & \multicolumn{4}{|c|}{Interaction layers} &  & \\
\hline
Focus & Participants & Domain & Strategies & \multicolumn{2}{|c|}{Styles} & Modalities & Robots, devices & Refs\\
\hline
 Creativity Support & Children & Storytelling & Stimulate human creativity & Demonstrate verbal creativity & Game-based & Speech, GUI & Jibo, tablet & \cite{ali_can_2019}\cite{ali2021social} \\
 & Children & Storytelling & Stimulate human creativity & Mirror/contrast user input & Open-ended & TUI & Robotic object (YOLO) & \cite{alves-oliveira_creativity_2020}\cite{alves-oliveira_guide_2019}\cite{alves-oliveira_software_2020} \\
 & Children & Storytelling & Stimulate human creativity & Promotional behavior & Open-ended & Speech, GUI & EMYS, tablet & \cite{elgarf_and_2022} \\
 & Children & Storytelling & Stimulate human creativity & Demonstrate verbal creativity & Open-ended & Speech, GUI & Furhat, tablet & \cite{elgarf_creativebot_2022}\cite{elgarf_once_2021} \\
 & Children & Storytelling & Stimulate human creativity & Scaffold creativity/reflection & Open-ended & TUI & Stuffed animal, physical tools & \cite{hubbard_child-robot_2021} \\
 & Children & Drawing & Stimulate human creativity & Embodied presence & Game-based & Speech, GUI & Jibo, tablet & \cite{ali_can_2019} \\
 & Children & Drawing & Stimulate human creativity & Demonstrate figural creativity & Game-based & Speech, GUI & Jibo, tablet & \cite{ali2021social} \\
 & Children & Construction & Stimulate human creativity & Scaffold creativity/reflection & Game-based & Speech, GUI & Jibo, tablet & \cite{ali2021social} \\
 & Adults & Storytelling & Stimulate human creativity & Demonstrate verbal creativity & Game-based & Speech, GUI & Robovie, display & \cite{kahn_creative_2014} \\
 & Adults & Drawing & Stimulate human creativity & Embodied presence & Game-based & Speech, GUI & Nao, tablet & \cite{alves-oliveira_sparking_2019} \\
 & Adults & Performance & Stimulate human creativity & Mirror/contrast user input & Open-ended & TUI & Robotic object, tablet & \cite{fabiano_designing_2017} \\
\hline
 Creative collaboration & Adults & Drawing & Non-verbal communication & Expressive robot movement & Open-ended & TUI, GUI & Cobot, physical tools & \cite{hinwood_proposed_2018} \\
 & Designers & Drawing & Conceptual sketching & Suggesting conceptual shifts & Open-ended & TUI & Drawbot, physical tools & \cite{lin_it_2020} \\
 & Adults & Drawing & Non-verbal communication & Expressive robot movement & Open-ended & TUI & Cobot, physical tools & \cite{pedersen_i_2020} \\
& Artists & Performance & Arts-led, process-led & Dialogue through improvisation & Open-ended & TUI, GUI & Cobot, tablet, dance floor & \cite{gomez_cubero_robot_2021} \\
& Children & Drawing & Human/machine learning & User-specific training data & Open-ended & TUI & Cobot, physical tools & \cite{twomey_three_2022} \\
\hline
Art Therapy & Adults & Painting & Responsive art approach & Express matching emotions & Open-ended & Speech, TUI & Baxter, BMI, physical tools & \cite{cooney_design_2018} \\
 & Adults & Painting & Balance contingency/artistry & Suggesting visual metaphors & Open-ended & Speech, TUI & Baxter, physical tools & \cite{cooney2019designing} \\
 & Adults & Painting & Connect to personal art & Personalized visual metaphors & Open-ended & Speech, TUI & Baxter, physical tools & \cite{cooney_robot_2021} \\
 & Children & Drawing & Personalization & Speech-assisted co-drawing & Open-ended & Speech, GUI & Cobot, tablet & \cite{shaik_co-creative_2021} \\
 \hline
 Artistic Work & Artist & Paint/perform & Human-machine symbiosis & Mimicry, memory a.o. & Open-ended & TUI, ambient & Cobots a.o., art installation & \cite{chung2022sketching} \\
 & Audience & Performance & Embodied Creative AI & Shared creative spaces & Open-ended & Ambient & Robotic objects, art installation & \cite{saunders_accomplice_2014} \\
 & Audience & Performance & Embodied Creative AI & Performance Body Mapping & Open-ended & Ambient & Robotic objects, art installation & \cite{saunders_performative_2018} \\
 & Audience & Paint/perform & Robotic art of audience input & Speech-to-AI-art & Open-ended & Speech, ambient & Kuka robot, art installation & \cite{sola_dream_2022} \\
 \hline
\end{tabular}}
\end{table*}

\section{RESULTS}\label{sec:results}

The FBS ontology~\cite{gero2014function} and Kantosalo's Interaction Framework for Human-Computer Co-Creativity~\cite{kantosalo2020modalities} were used for the analysis of reviewed articles. An overview is presented in Table~\ref{tab:review}.

\subsection{Function}

When looking at the reviewed articles presented in Table~\ref{tab:review}, we can distinguish four categories based on their research focus: 1) Creativity Support 2) Creative Collaboration 3) Art Therapy, and 4) Artistic Work. \emph{Creativity Support} forms the largest group, with studies investigating factors of social robot behavior that affect human creativity. Studies of \emph{Creative Collaboration} explored the interaction dynamics, and how the process of collaboration can be facilitated. In the category of \emph{Art Therapy}, the focus was on specific therapeutic requirements, and how to design responsive systems for affective and assistive collaborative painting and drawing. In the category of \emph{Artistic Work}, the focus is on creative encounters between humans and machines, and studies are carried out in the context of the researchers' own artistic practice, mostly in performances involving the audience. Regarding participants and target groups, we found that studies on \emph{Creativity Support} involved mostly children, while adults and professional artists and designers participated in the \emph{Creative Collaboration} studies. In the category of \emph{Artistic Work}, the artists themselves also played an important role, as well as the audience.

\subsection{Behavior: Strategies, Styles, Modalities}

All studies in the category of \emph{Creativity Support} propose the strategy of stimulating human creativity, through various social behaviors of a robot. Creativity demonstration was used to stimulate creativity with children (n=4) and with adults (n=1). The robot demonstrated verbal creativity in storytelling applications and figural creativity in a drawing game. It was found that creativity demonstrations and scaffolding (e.g. asking questions, prompting, and suggestions), as well as the promotion behavior of the robot can contribute to higher levels of human creativity. When mirroring or contrasting robot movements were congruent with user input, this positively affected creativity \cite{alves-oliveira_creativity_2020}\cite{fabiano_designing_2017}. The studies mostly compared conditions of robot behavior, using pre-defined, validated content. For example, in multiple studies the robot demonstrated creativity by selecting pre-defined suggestions with a validated creativity score, dependent on the condition~\cite{ali_can_2019}\cite{ali2021social}\cite{kahn_creative_2014}.

In the category of \emph{Creative Collaboration}, two studies explored expressive robot movements to improve non-verbal communication \cite{hinwood_proposed_2018}\cite{pedersen_i_2020}. In the context of collaborative drawing, the effects of direct versus indirect motion paths on collaborative interaction were compared, but the results were inconclusive. The researchers recommend further in-the-field experiments, combining qualitative and quantitative methodologies. An arts-led, process-led approach is proposed by Gomez Cubero et al.~\cite{gomez_cubero_robot_2021}, to explore how co-creativity emerges through human-robot dialogue and improvisation. The researchers developed custom tools to support collaborative drawing with an industrial robot and put these into practice. In a study involving designers, a mobile robot was introduced for collaborative sketching and generating ideas through `conceptual shifts'~\cite{lin_it_2020}. Using the Sketch-RNN model and the {Google Quick, Draw!} API, input sketches were mapped to suggestions with visual and semantic similarity. Results showed that the mobile, embodied agent performed better in provoking exploratory thinking and collaborative ideation, compared to a web-based agent. The alignment of human, robot, and machine learning is suggested by Twomey~\cite{twomey_three_2022} in a study where the robotic system is trained on audience-specific content in the form of children's drawings.

With \emph{Art Therapy}, the focus is on investigating how a robot can learn to understand and adapt to the creative and emotional expressions of a human interaction partner. Cooney \& Menezes~\cite{cooney_design_2018} propose to generate responsive art for emotion regulation, through robot expressions of either matching or positive emotions. Using wireless electroencephalography (EEG), brain signals were captured and classified based on Russell's valence/arousal model, and then translated into visual features for paintings. Affective image databases were used to train the system and Deep Convolutional Generative Adversarial Networks (DC-GANs) for synthesizing compositions. To balance contingency and artistry, Cooney \& Berck~\cite{cooney2019designing} made use of visual metaphors that are responsive to perceived emotions. In a follow-up study, Cooney~\cite{cooney_robot_2021} proposes metaphors that connect more to emotional artistic expressions. Personalization was also facilitated, through the robot's open questions, letting users add their own tags to describe the content. Another implementation of personalization is suggested by Shaik et al.~\cite{shaik_co-creative_2021}, by letting the system adapt sketches based on verbal feedback and explicit directions from users, which were disabled children.

\begin{figure*}[htb]
\centering
\includegraphics[width=\textwidth]{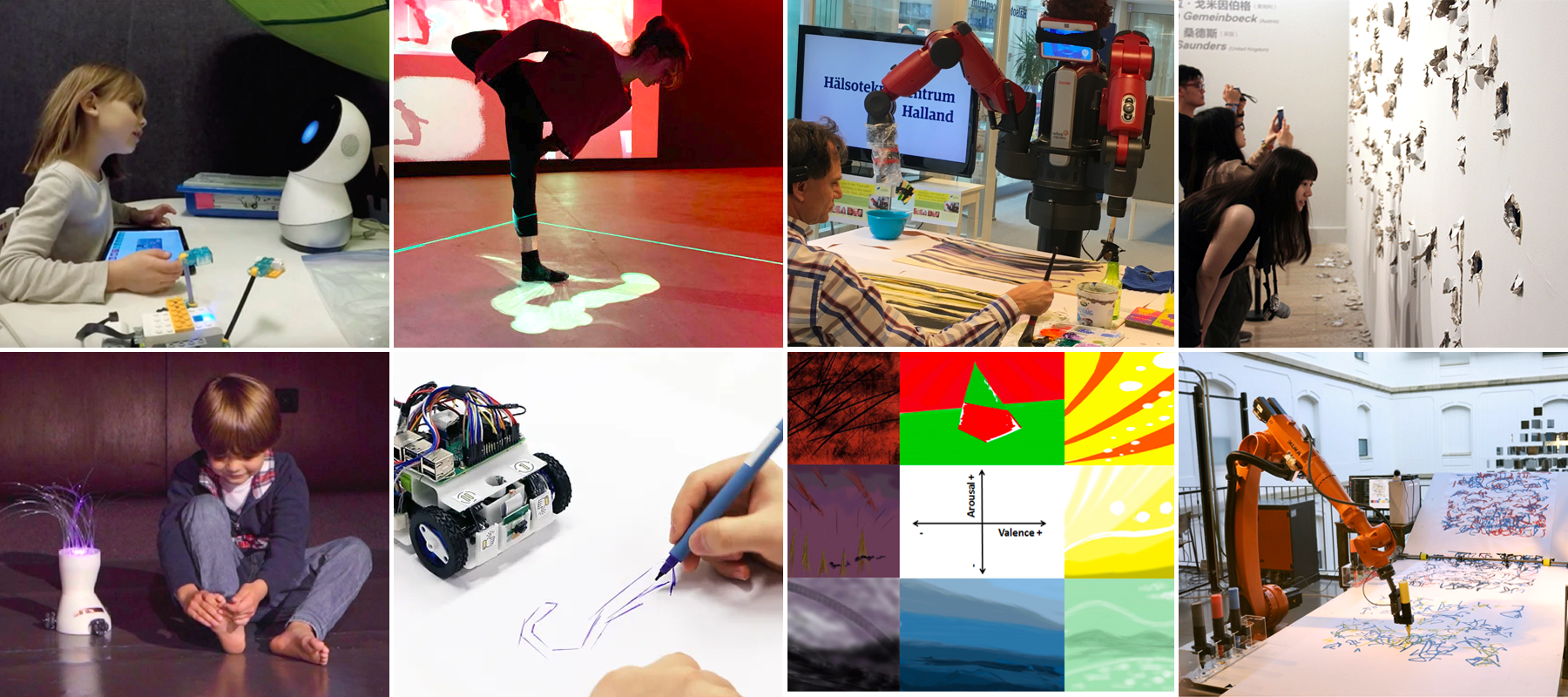}
\caption{Per column: 1) Jibo, scaffolding creativity in a construction task \cite{ali2021social}; YOLO, a robot toy for storytelling \cite{alves-oliveira_creativity_2020}; 2) The robot is present, an arts-led, process-led approach for investigating human-robot dialogues in improvisation \cite{gomez_cubero_robot_2021}; Cobbie, a drawbot for conceptual sketching with designers \cite{lin_it_2020}; 3) Baxter robot used in Art Therapy; The valence/arousal model for expressing matching emotions; 4) Accomplice - Creative robotics and embodied computational creativity \cite{saunders_accomplice_2014}; Dream Painter - Bridging Audience Interaction, Robotics, and Creative AI \cite{sola_dream_2022}}
\label{fig:structure}
\end{figure*}

In the category of \emph{Artistic Work}, Sougwen Chung~\cite{chung2022sketching} explores human-machine symbiosis, studying concepts of mimicry, memory, collectivity, and spectrality. For example, mimicry is explored with the robot mimicking the artists drawing gestures and for memory, the machine learns the artist's drawing style, with neural nets trained on the artists drawing collection. Saunders \& Gemeinboeck~\cite{saunders_accomplice_2014} investigate how embodied, creative AI can act as a performer by embedding a group of autonomous robots into the walls of a gallery. The robots are programmed as curious agents, driven to explore their world. By punching on the walls and making holes, they make changes to the environment, communicate their presence, and involve the audience. In another study by Saunders \& Gemeinboeck~\cite{saunders_performative_2018}, professional dancers and non-humanlike robots were brought together in co-embodied explorations of forms and movements. Sola et al.~\cite{sola_dream_2022} suggest speech-to-AI-art transformations and created an interface that allowed the audience to tell a co-creative system about their dreams. Based on prompts from the input text, the AI-system generated a drawing through latent space navigation using the CLIP model ~\cite{radford2021learning}. The industrial robot arm captured the audience's stories in a collective painting that hang down into the atrium of a museum as a cascade of dreams.

When looking at game-based versus open-ended interaction styles, the studies that compared robot behavior in different conditions generally used game-based interaction, e.g., in drawing and storytelling games. This facilitated experimental control when comparing and measuring the effects of social robot behavior. In other studies, open-ended forms of interaction were used, which allows for investigating processes and dynamics, and contributes to ecological validity when researching creative collaboration.

What stands out when looking at the interaction modalities used in the four categories, is that speech is mostly used in the categories \emph{Creativity Support} and \emph{Art Therapy}. Robot speech is used for demonstrating verbal creativity, and for scaffolding creativity, and prompting creative reflection \cite{ali2021social}\cite{hubbard_child-robot_2021}. For art therapy robots, speech is found to be useful as well. It allows complex information to be conveyed, in a familiar and intuitive way, without requiring a person to look away from art-making and possibly lose concentration \cite{cooney_design_2018}. Speech also enables users to give explicit feedback or directions or ask for assistance, as suggested by Shaik et al.~\cite{shaik_co-creative_2021}. In the categories \emph{Creative Collaboration} and \emph{Artistic Work}, human-machine dialogues are more often based on non-verbal communication such as expressive movement, or through the work itself. With \emph{Artistic Work}, we see most examples of ambient interfaces, when exploring new forms of human-machine encounters in a spatial setting (Fig.~\ref{fig:structure}). 

\subsection{Structure}

Different types of robots and embodiments were used in the reviewed studies (Fig.~\ref{fig:structure}). Human-like robots were used in almost all studies in using a \textit{Stimulating creativity} strategy, combined with a tablet or a computer screen. With the drawing activities in this category, the robot (Jibo, Nao) was not drawing physically, but virtually on the tablet, with separate canvases on the same tablet \cite{ali_can_2019}\cite{ali_social_2021}\cite{alves-oliveira_sparking_2019}. In studies employing \textit{game-based} interaction styles, screens were used to present the game world. Alves-Oliveira et al.~\cite{alves-oliveira_creativity_2020}\cite{alves-oliveira_guide_2019}\cite{alves-oliveira_software_2020} used a non-anthropomorphic robot object to stimulate creativity in children; YOLO serves as a toy character in a storytelling game. The robot interacts through lights, colors, and movements, while the shape of the robot set realistic expectations for the robot's capabilities. In the category of \emph{Creative Collaboration}, collaborative robot (cobot) arms were mostly used, together with physical drawing tools. Physical drawing tools were also used with art therapy robots. The Baxter robot used for art therapy \cite{cooney_design_2018} can be considered a human-like cobot, with two arms and a screen that can display a face and facial expressions, facilitating non-verbal social communication. 
With Artistic Work, robot arms were used next to custom-made robotic objects, mostly in multi-agent settings. The stage is shared between humans and robots, mostly in performances. In an art installation by Sola et al.~\cite{sola_dream_2022}, the industrial robot arm is behind glass, while the audience can communicate with the robot through a speech interface. In Accomplice, Saunders \& Gemeinboeck~\cite{saunders_accomplice_2014} install robots in their own space behind a wall in a gallery, which they breaking through as they use the wall as their canvas. Saunders \& Gemeinboeck~\cite{saunders_performative_2018} used robotic cubes to explore how human and non-human forms of embodiment can be mapped through movement, and how non-humanlike robotic objects can be perceived as affective agents.

\begin{figure*}
\centering
\includegraphics[width=0.8\linewidth]{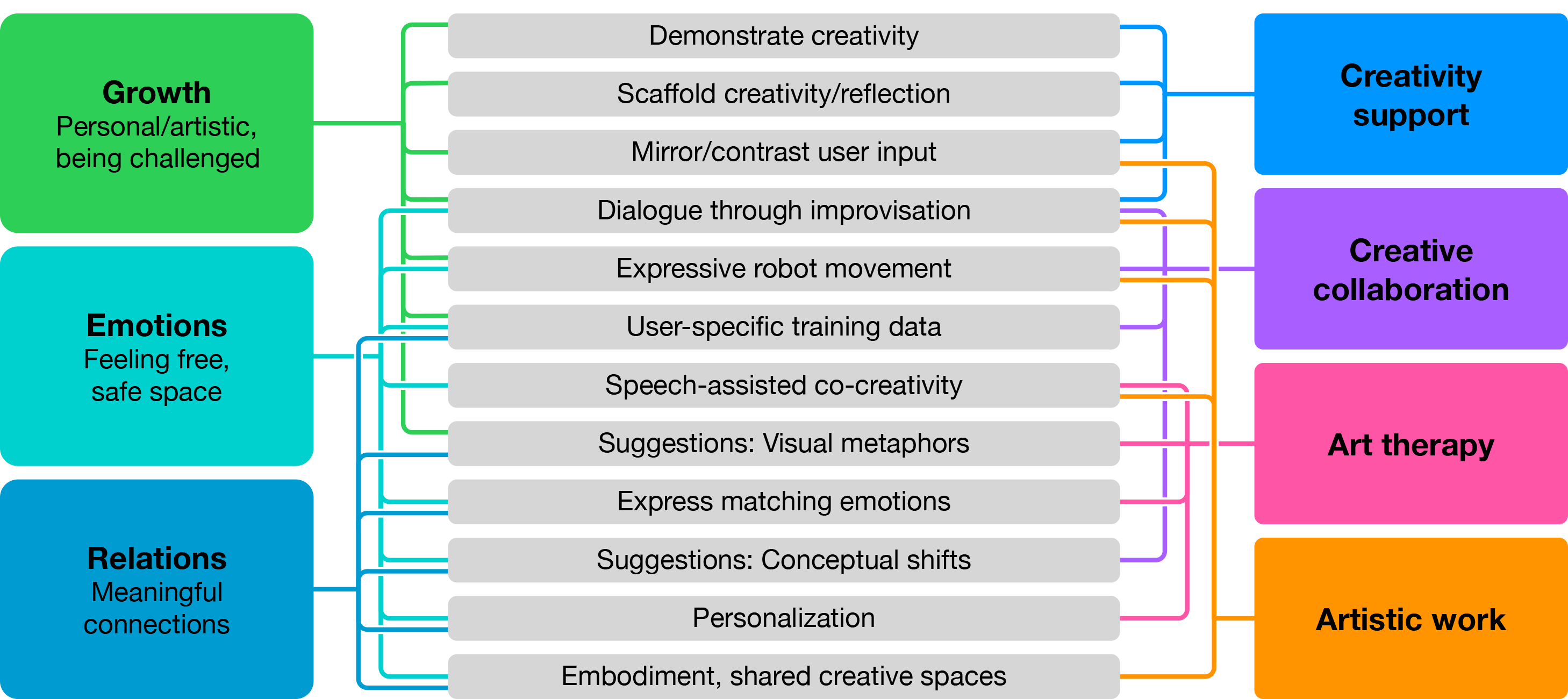}
\caption{Values that older adults attributed to their creative activities in a nationwide Dutch study \cite{groot2021value}, connected to robot behaviors suggested in the reviewed studies, with the corresponding categories (Table~\ref{tab:review}). These values (e.g. meaningful connections) were attributed in the context of human-human interactions. We suggest to investigate if and how human-robot co-creative interactions can be valuable to older adults as well. We propose this mapping of values and behaviors as part of our research agenda (Section~\ref{sec:agenda}).} 
\label{fig:connections}
\end{figure*}

\section{DISCUSSION}\label{sec:discussion}
This review has several limitations. An important limitation is the fact that the review selection was carried out by a single reviewer, due to time constraints. In addition, there are opportunities to further work out the analysis, for example to explore how machine learning techniques, as part of the structure of a system, align with behavior and function. It turned out that our search results did not include any studies involving the elderly. This shows that future research on HRCC with older adults is important. However, it is also a limitation, as we cannot learn from previous findings in HRCC with the target group. We are planning an extended version of this review, with a search query that does include the target group, f.e., looking at gerontechnology for creativity support. This allows us to involve more reviewers and to address aspects that have been underexposed so far. 

\section{CONCLUSION}\label{sec:conclusion}
Selected articles were structured using the FBS ontology \cite{gero2014function} and the Interaction Framework for Human-Computer Co-creativity \cite{kantosalo2020modalities}. The search and selection process (see Section~\ref{sec:method}) resulted in a heterogeneous set of studies, describing robotic systems with various functions, behaviors, and structures. 

 \subsection{Function}
 Studies in the categories of \emph{Creativity Support} and \emph{Art Therapy} take a developmental perspective, with the goal to a) stimulate human creativity and b) support art therapy through responsiveness and personalization. Studies in the categories \emph{Creative Collaboration} and \emph{Artistic Work} take an ecological perspective, investigating how creativity emerges through interaction. This is a more process-led approach, involving end users and taking into account the social and material environment. As set out in Section~\ref{sec:background}, both ecological and developmental perspectives align with values that older adults attribute to their creative experience, and must be taken into account when defining the functions of HRCC for older adults. An important finding regarding participants is that older adults did not engage in any of the reviewed studies. While Cooney \& Menezes~ \cite{cooney_design_2018} thank older adults in their acknowledgments for providing input, they evaluated their system with younger adults. It is not clear why older adults have not yet been involved in HRCC research. Robots and AI-generated content offer opportunities that can be beneficial for this specific target group, which is growing worldwide, and there are specific needs and wishes to be taken into account. That is why we are making a case for investigating HRCC for, and with, the target group of older adults.

\subsection{Behavior}
Evidence shows that robots are capable of demonstrating creativity, and that this social behavior can be designed to stimulate human creativity. Other social behaviors are found to be effective as well, such as mirroring and contrasting user input to promote divergent and convergent thinking. Studies on robots in \emph{Art Therapy} provide valuable insights into the importance of recognizing, modeling, and synthesizing emotions in drawings and paintings. Here, the emphasis is on tailoring and balancing content to user needs e.g., using personalized visual metaphors. These ideas on how an art therapy robot could behave as a `third hand' also inform future research in HRCC for older adults. Studies in the category of \emph{Creativity Support} often used games to structure the interaction, which contributes to experimental control when measuring the effects of robot behavior. However, the majority of studies used open-ended forms of interaction, investigating how dialogues and collaborations develop.  

The modality of speech is considered an important channel for transparency and effective communication, promoting autonomy and control. This is emphasized in the categories \emph{Creativity Support} and \emph{Art Therapy}. Robot speech is used, e.g., to demonstrate verbal creativity, scaffold creativity, and promote creative reflection. User speech input is suggested as a means for explicit feedback, requesting assistance, and personalizing suggested content. The research projects in the category of \emph{Artistic Work} place creative robots in spatial settings, sometimes with multiple agents, and letting artists and audiences contribute to a physical shared space that fosters creativity. Both speech and embodied, spatial interactions are of interest for HRCC with older adults, to contribute to an environment where people feel free and safe.

\subsection{Structure}
Results show that in the categories \emph{Creativity Support} and \emph{Art Therapy}, mostly human-like robots were used. The robot YOLO is an exception, an abstract robotic object that serves as a toy, while the shape of the robot sets realistic expectations for the robot's capabilities \cite{alves-oliveira_guide_2019}. A shared stage for humans and robots, as explored in studies on \emph{Artistic Work}, could be interesting for older adults as well, when designed as an environment fostering creativity, and where people feel free and safe.

\section{RESEARCH AGENDA}\label{sec:agenda}
We propose a participatory, value-sensitive design approach for investigating HRCC with older adults. Older adults must be involved throughout the entire process, in identifying opportunities and requirements, developing HRCC activities, and testing hypotheses in both controlled experiments and in-the-field settings. When investigating the design of the system, we propose considering of the following aspects, aligned with the the FBS framework:
\begin{boxed}
\\
\textbf{Function:} Consider both ecological and developmental perspectives on creativity when defining functional requirements for the target group. \\
\\
\textbf{Behavior:} Align values that older adults attribute to creative activities with the opportunities of HRCC (Fig.~\ref{fig:connections}) to investigate how:
\begin{enumerate}
    {\item A robot's social behavior can support and enhance creative experiences for older adults;
    \item AI-generated content can be tailored and responsive to specific needs and desires; and,
    \item Intuitive dialogues (verbal, non-verbal, through artifacts) can support co-creativity.}\vspace*{-\baselineskip}
\end{enumerate} \\
\\
\textbf{Structure:} Investigate what types of robot and devices fit best and provide opportunities for:
\begin{enumerate}
    {\item Social interaction with older adults;
    \item Creative support and exploration; and,
    \item Shared creative experiences and spaces where older adults feel free and safe.}\vspace*{-\baselineskip}
\end{enumerate}
\\
\end{boxed}


\addtolength{\textheight}{-3.5cm}   




\section*{ACKNOWLEDGMENT}

This publication is part of the project `Social robotics and generative AI to support and enhance creative experiences for older adults', with project number 023.019.021 of the research program Doctoral Grant for Teachers which is financed by the Dutch Research Council (NWO). 


\bibliographystyle{ieeetr}
\bibliography{articles}

\begin{thebibliography}{10}

\bibitem{united2022world}
{United Nations}, ``World population prospects 2022: Summary of results,''
  2022.

\bibitem{fancourt2019evidence}
D.~Fancourt and S.~Finn, {\em What is the evidence on the role of the arts in
  improving health and well-being? A scoping review}.
\newblock World Health Organization. Regional Office for Europe, 2019.

\bibitem{cohen2005mature}
G.~D. Cohen, {\em The Mature Mind: The Positive Power of the Aging Brain}.
\newblock {Basic Books} ({AZ}), 2005.

\bibitem{creech2020creativity}
A.~Creech, K.~Larouche, M.~Generale, and D.~Fortier, ``Creativity, music, and
  quality of later life: A systematic review,'' {\em Psychology of Music},
  p.~0305735620948114, 2020.

\bibitem{zeilig2019co}
H.~Zeilig, V.~Tischler, M.~van~der Byl~Williams, J.~West, and S.~Strohmaier,
  ``Co-creativity, well-being and agency: A case study analysis of a
  co-creative arts group for people with dementia,'' {\em Journal of Aging
  Studies}, vol.~49, pp.~16--24, 2019.

\bibitem{broekens-heerink-rosendal:2009assistive}
J.~Broekens, M.~Heerink, and H.~Rosendal, ``Assistive social robots in elderly
  care: A review,'' {\em Gerontechnology}, vol.~8, no.~2, pp.~94--103, 2009.

\bibitem{allouch2020social}
S.~B. Allouch and L.~van Velsen, ``Social robots for elderly care: An inventory
  of promising use cases and business models.,'' in {\em MIE}, pp.~1046--1050,
  2020.

\bibitem{lubart_creativity_2021}
T.~Lubart, D.~Esposito, A.~Gubenko, and C.~Houssemand, ``Creativity in humans,
  robots, humbots,'' {\em Creativity}, vol.~8, no.~1, pp.~23--37, 2021.

\bibitem{deterding2017mixed}
S.~Deterding, J.~Hook, R.~Fiebrink, M.~Gillies, J.~Gow, M.~Akten, G.~Smith,
  A.~Liapis, and K.~Compton, ``Mixed-initiative creative interfaces,'' in {\em
  Proceedings of the 2017 {CHI} Conference Extended Abstracts on Human Factors
  in Computing Systems}, pp.~628--635, 2017.

\bibitem{runco-jaeger:2012standard}
M.~A. Runco and G.~J. Jaeger, ``The standard definition of creativity,'' {\em
  Creativity Research Journal}, vol.~24, no.~1, pp.~92--96, 2012.

\bibitem{gluaveanu2013rewriting}
V.~P. Gl{\u{a}}veanu, ``Rewriting the language of creativity: The {Five A's}
  framework,'' {\em Review of General Psychology}, vol.~17, no.~1, pp.~69--81,
  2013.

\bibitem{kaufman2009beyond}
J.~C. Kaufman and R.~A. Beghetto, ``Beyond big and little: The {Four C} model
  of creativity,'' {\em Review of General Psychology}, vol.~13, no.~1,
  pp.~1--12, 2009.

\bibitem{groot2021value}
B.~Groot, L.~de~Kock, Y.~Liu, C.~Dedding, J.~Schrijver, T.~Teunissen, M.~van
  Hartingsveldt, J.~Menderink, Y.~Lengams, J.~Lindenberg, {\em et~al.}, ``The
  value of active arts engagement on health and well-being of older adults: A
  nation-wide participatory study,'' {\em International Journal of
  Environmental Research and Public Health}, vol.~18, no.~15, p.~8222, 2021.

\bibitem{liu2022participatory}
Y.~Liu, B.~Groot, L.~de~Kock, T.~Abma, and C.~Dedding, ``How participatory arts
  can contribute to {Dutch} older adults' wellbeing -- revisiting a taxonomy of
  arts interventions for people with dementia,'' {\em Arts \& Health},
  pp.~1--16, 2022.

\bibitem{pease-et-al.:2022roadmap}
A.~Pease, M.~Ackerman, N.~Pease, and B.~McFadden, ``A roadmap for therapeutic
  computational creativity,'' in {\em Proceedings of the 13th International
  Conference on Computational Creativity} (M.~M. Hedblom, A.~A. Kantosalo,
  R.~Confalonieri, O.~Kutz, and T.~Veale, eds.), pp.~261--270, Association for
  Computational Creativity {(ACC)}, June 27 -- July 1 2022.

\bibitem{compton2015casual}
K.~Compton and M.~Mateas, ``Casual creators,'' in {\em Proceedings of the Sixth
  International Conference on Computational Creativity, {ICCC 2015}},
  pp.~228--235, 2015.

\bibitem{kramer1986art}
E.~Kramer, ``The art therapist's third hand: Reflections on art, art therapy,
  and society at large,'' {\em American Journal of Art Therapy}, 1986.

\bibitem{pease2022roadmap}
A.~Pease, M.~Ackerman, N.~Pease, and B.~McFadden, ``A roadmap for therapeutic
  computational creativity,'' in {\em Proceedings of the 13th International
  Conference on Computational Creativity ({ICCC'22})} (M.~M. Hedblom, A.~A.
  Kantosalo, R.~Confalonieri, O.~Kutz, and T.~Veale, eds.), pp.~261--270,
  Association for Computational Creativity {(ACC)}, June 27 -- July 1 2022.

\bibitem{gemeinboeck2016performance}
P.~Gemeinboeck and R.~Saunders, ``The performance of creative machines,'' in
  {\em Cultural Robotics: First International Workshop, {CR 2015}, Held as Part
  of {IEEE RO-MAN 2015}, {Kobe}, {Japan}, {August} 31, 2015. Revised Selected
  Papers 1}, pp.~159--172, Springer, 2016.

\bibitem{bray_applying_2016}
L.~Bray and O.~Bown, ``Applying core interaction design principles to
  computational creativity,'' in {\em Proceedings of the 7th International
  Conference on Computational Creativity, {ICCC} 2016}, pp.~93--97, 2016.

\bibitem{bown2020speculative}
O.~Bown, K.~Grace, L.~Bray, and D.~Ventura, ``A speculative exploration of the
  role of dialogue in human-computer co-creation,'' in {\em Proceedings of the
  11th International Conference on Computational Creativity, {ICCC} 2020},
  pp.~25--32, June 29 -- July, 2020 2020.

\bibitem{ouzzani2016rayyan}
M.~Ouzzani, H.~Hammady, Z.~Fedorowicz, and A.~Elmagarmid, ``Rayyan -- a web and
  mobile app for systematic reviews,'' {\em Systematic Reviews}, vol.~5,
  pp.~1--10, 2016.

\bibitem{moher2009preferred}
D.~Moher, A.~Liberati, J.~Tetzlaff, D.~G. Altman, and {The PRISMA Group},
  ``Preferred reporting items for systematic reviews and meta-analyses: The
  {PRISMA} statement,'' {\em Annals of Internal Medicine}, vol.~151, no.~4,
  pp.~264--269, 2009.
\newblock PMID: 19622511.

\bibitem{gero2014function}
J.~S. Gero and U.~Kannengiesser, ``The {Function-Behaviour-Structure} ontology
  of design,'' in {\em An Anthology of Theories and Models of Design:
  Philosophy, Approaches and Empirical Explorations} (A.~Chakrabarti and
  L.~T.~M. Blessing, eds.), pp.~263--283, London: Springer, 2014.

\bibitem{kantosalo2020modalities}
A.~Kantosalo, P.~T. Ravikumar, K.~Grace, and T.~Takala, ``Modalities, styles
  and strategies: An interaction framework for human-computer co-creativity.,''
  in {\em Proceedings of the 11th International Conference on Computational
  Creativity, {ICCC} 2020}, pp.~57--64, June 29 -- July, 2020 2020.

\bibitem{ali_can_2019}
S.~Ali, T.~Moroso, and C.~Breazeal, ``Can children learn creativity from a
  social robot?,'' in {\em Proceedings of the 2019 Creativity and Cognition
  ({C\&C 2019})}, (New York, NY, USA), pp.~359--368, Association for Computing
  Machinery, 2019.

\bibitem{ali2021social}
S.~Ali, N.~Devasia, H.~W. Park, and C.~Breazeal, ``Social robots as creativity
  eliciting agents,'' {\em Frontiers in Robotics and AI}, vol.~8, 2021.

\bibitem{alves-oliveira_creativity_2020}
P.~Alves-Oliveira, P.~Arriaga, M.~Cronin, and A.~Paiva, ``Creativity encounters
  between children and robots,'' in {\em {ACM/IEEE} International Conference on
  Human-Robot Interaction ({HRI})}, pp.~379--388, 2020.

\bibitem{alves-oliveira_guide_2019}
P.~Alves-Oliveira, P.~Arriaga, A.~Paiva, and G.~Hoffman, ``Guide to build
  {YOLO}, a creativity-stimulating robot for children,'' {\em HardwareX},
  vol.~6, 2019.

\bibitem{alves-oliveira_software_2020}
P.~Alves-Oliveira, S.~Gomes, A.~Chandak, P.~Arriaga, G.~Hoffman, and A.~Paiva,
  ``Software architecture for {YOLO}, a creativity-stimulating robot,'' {\em
  SoftwareX}, vol.~11, 2020.

\bibitem{elgarf_and_2022}
M.~Elgarf, N.~Calvo-Barajas, P.~Alves-Oliveira, G.~Perugia, G.~Castellano,
  C.~Peters, and A.~Paiva, ``'and then what happens?' promoting children's
  verbal creativity using a robot,'' in {\em {ACM/IEEE} International
  Conference on Human-Robot Interaction ({HRI} 2022)}, pp.~71--79, 2022.

\bibitem{elgarf_creativebot_2022}
M.~Elgarf, S.~Zojaji, G.~Skantze, and C.~Peters, ``{CreativeBot}: A creative
  storyteller robot to stimulate creativity in children,'' in {\em Proceedings
  of the 2022 International Conference on Multimodal Interaction, {ICMI} '22},
  (New York, NY, USA), p.~540–548, Association for Computing Machinery, 2022.

\bibitem{elgarf_once_2021}
M.~Elgarf, G.~Skantze, and C.~Peters, ``Once upon a story: Can a creative
  storyteller robot stimulate creativity in children?,'' in {\em Proceedings of
  the 21st {ACM} International Conference on Intelligent Virtual Agents, {IVA}
  2021}, pp.~60--67, 2021.

\bibitem{hubbard_child-robot_2021}
L.~J. Hubbard, Y.~Chen, E.~Colunga, P.~Kim, and T.~Yeh, ``Child-robot
  interaction to integrate reflective storytelling into creative play,'' in
  {\em Proceedings of the 2021 Creativity and Cognition ({C\&C 2021})}, (New
  York, NY, USA), Association for Computing Machinery, 2021.

\bibitem{kahn_creative_2014}
P.~H. Kahn, T.~Kanda, H.~Ishiguro, S.~Shen, H.~E. Gary, and J.~H. Ruckert,
  ``Creative collaboration with a social robot,'' in {\em Proceedings of the
  2014 {ACM} International Joint Conference on Pervasive and Ubiquitous
  Computing}, {UbiComp} '14, (New York, NY, USA), p.~99–103, Association for
  Computing Machinery, 2014.

\bibitem{alves-oliveira_sparking_2019}
P.~Alves-Oliveira, S.~Tulli, P.~Wilken, R.~Merhej, J.~Gandum, and A.~Paiva,
  ``Sparking creativity with robots: A design perspective,'' in {\em 14th
  Annual {ACM/IEEE} International Conference on Human Robot Interaction
  ({HRI})}, Jan. 2019.

\bibitem{fabiano_designing_2017}
F.~Fabiano, H.~Pelikan, J.~Pingen, J.~Zissoldt, A.~Catala, and M.~Theune,
  ``Designing a co-creative dancing robotic tablet,'' in {\em {CEUR} Workshop
  Proceedings}, vol.~2160, 2017.

\bibitem{hinwood_proposed_2018}
D.~Hinwood, J.~Ireland, E.~Jochum, and D.~Herath, ``A proposed {Wizard of OZ}
  architecture for a human-robot collaborative drawing task,'' {\em Lecture
  Notes in Computer Science}, vol.~11357, pp.~35--44, 2018.

\bibitem{lin_it_2020}
Y.~Lin, J.~Guo, Y.~Chen, C.~Yao, and F.~Ying, ``It is your turn: Collaborative
  ideation with a co-creativ robot through sketch,'' in {\em Conference on
  Human Factors in Computing Systems}, 2020.

\bibitem{pedersen_i_2020}
J.~Pedersen, K.~Christensen, D.~Herath, and E.~Jochum, ``I like the way you
  move: A mixed-methods approach for studying the effects of robot motion on
  collaborative human robot interaction,'' {\em Lecture Notes in Computer
  Science}, vol.~12483, pp.~73--84, 2020.

\bibitem{gomez_cubero_robot_2021}
C.~Gomez~Cubero, M.~Pekarik, V.~Rizzo, and E.~Jochum, ``The robot is present:
  Creative approaches for artistic expression with robots,'' {\em Frontiers in
  Robotics and AI}, vol.~8, 2021.

\bibitem{twomey_three_2022}
R.~Twomey, ``Three stage drawing transfer: Collaborative drawing between a
  {Generative Adversarial Network}, co-robotic arm, and five-year-old child,''
  {\em Proceedings of the {ACM} on Computer Graphics and Interactive
  Techniques}, vol.~5, no.~4, 2022.

\bibitem{cooney_design_2018}
M.~Cooney and M.~Menezes, ``Design for an art therapy robot: An explorative
  review of the theoretical foundations for engaging in emotional and creative
  painting with a robot,'' {\em Multimodal Technologies and Interaction},
  vol.~2, no.~3, 2018.

\bibitem{cooney2019designing}
M.~Cooney and P.~Berck, ``Designing a robot which paints with a human: Visual
  metaphors to convey contingency and artistry,'' in {\em {ICRA-X} Robots Art
  Program at {IEEE} International Conference on Robotics and Automation
  ({ICRA}), Montreal QC, Canada}, p.~2, 2019.

\bibitem{cooney_robot_2021}
M.~Cooney, ``Robot art, in the eye of the beholder?: Personalized metaphors
  facilitate communication of emotions and creativity,'' {\em Frontiers in
  Robotics and {AI}}, vol.~8, 2021.

\bibitem{shaik_co-creative_2021}
S.~Shaik, V.~Srinivasan, Y.~Peng, M.~Lee, and N.~Davis, ``Co-creative robotic
  arm for differently-abled kids: Speech, sketch inputs and external feedbacks
  for multiple drawings,'' {\em Advances in Intelligent Systems and Computing},
  vol.~1290, pp.~998--1007, 2021.

\bibitem{chung2022sketching}
S.~Chung, ``Sketching symbiosis: Towards the development of relational
  systems,'' in {\em The Language of Creative {AI}: Practices, Aesthetics and
  Structures}, pp.~259--276, Springer, 2022.

\bibitem{saunders_accomplice_2014}
R.~Saunders and P.~Gemeinboeck, ``{Accomplice}: Creative robotics and embodied
  computational creativity,'' in {\em Proceedings of {AISB 2014} -- 50th Annual
  Convention of the {SSAISB}}, (London, UK), Society for the Study of
  Artificial Intelligence and the Simulation of Behaviour, 2014.

\bibitem{saunders_performative_2018}
R.~Saunders and P.~Gemeinboeck, ``{Performative Body Mapping} for designing
  expressive robots,'' in {\em Proceedings of the 9th International Conference
  on Computational Creativity, {ICCC 2018}}, pp.~280--287, 2018.

\bibitem{sola_dream_2022}
M.~Canet~Sola and V.~Guljajeva, ``{Dream Painter}: Exploring creative
  possibilities of {AI}-aided speech-to-image synthesis in the interactive art
  context,'' {\em Proceedings of the {ACM} on Computer Graphics and Interactive
  Techniques}, vol.~5, no.~4, 2022.

\bibitem{radford2021learning}
A.~Radford, J.~W. Kim, C.~Hallacy, A.~Ramesh, G.~Goh, S.~Agarwal, G.~Sastry,
  A.~Askell, P.~Mishkin, J.~Clark, G.~Krueger, and I.~Sutskever, ``Learning
  transferable visual models from natural language supervision,'' 2021.

\bibitem{ali_social_2021}
S.~Ali, H.~Park, and C.~Breazeal, ``A social robot's influence on children's
  figural creativity during gameplay,'' {\em International Journal of
  Child-Computer Interaction}, vol.~28, 2021.

\end{thebibliography}

\end{document}